\documentclass[a4paper,USenglish]{lipics-v2018}

\usepackage{accents}
\usepackage{amsmath}
\usepackage{amssymb}
\usepackage{cite}
\usepackage{float}
\usepackage{wasysym}
\usepackage[utf8]{inputenc}
\usepackage{microtype}
\usepackage{pgfplots}
\usepackage{tabto}
\usepackage{tikz}
\usetikzlibrary{calc}
\usetikzlibrary{patterns}
\usepackage[disable]{todonotes}
\usepackage{mathtools}

\pgfplotsset{compat=1.14}
\nolinenumbers

\title{Exact Algorithms for the Maximum Planar Subgraph Problem: New Models and Experiments}
\titlerunning{Exact Algorithms for Maximum Planar Subgraph}

\EventEditors{Gianlorenzo D'Angelo}
\EventNoEds{1}
\EventLongTitle{17th International Symposium on Experimental Algorithms (SEA 2018)}
\EventShortTitle{SEA 2018}
\EventAcronym{SEA}
\EventYear{2018}
\EventDate{June 27--29, 2017}
\EventLocation{L'Aquila, Italy}
\EventLogo{}
\SeriesVolume{103}
\ArticleNo{22}
\author{Markus Chimani}{Theoretical Computer Science, Osnabrück University, Germany}{markus.chimani@uni-osnabrueck.de}{0000-0002-4681-5550}{Supported by the German Research Foundation (DFG) project CH 897/2-1.}
\author{Ivo Hedtke}{Data Strategy \& Analytics, Schenker AG, Essen, Germany}{ivo.hedtke@dbschenker.com}{0000-0003-0335-7825}{}
\author{Tilo Wiedera$^1$}{Theoretical Computer Science, Osnabrück University, Germany}{tilo.wiedera@uni-osnabrueck.de}{0000-0002-5923-4114}{}
\authorrunning{M. Chimani, I. Hedtke, and T. Wiedera}
\Copyright{Markus Chimani, Ivo Hedtke, and Tilo Wiedera}

\subjclass{}


\newcommand{\pd}{\text{p.d.}}
\newcommand{\cpp}{\texttt{C++}}
\newcommand{\subsetOrder}[2]{{#1}^{\{#2\}}}
\newcommand{\crossOrder}[2]{{#1}^{(#2)}}
\newcommand{\ctarget}[1]{\ensuremath{\accentset{\raisebox{.045em}{\scalebox{.3}\newmoon}}{#1}}}
\newcommand{\csource}[1]{\ensuremath{\accentset{\circ}{#1}}}
\newcommand{\crev}[1]{\mathrm{rev}(#1)}
\newcommand{\dedge}[1]{\mathrm{e}(#1)}
\newcommand{\noinci}[1]{V^{#1}}
\newcommand{\neighbors}[1]{\mathrm{N}(#1)}
\newcommand{\din}[1]{\mathrm{\delta}^-(#1)}
\newcommand{\dout}[1]{\mathrm{\delta}^+(#1)}
\newcommand{\tsum}{\textstyle\sum}
\newcommand{\optsymbol}{\ensuremath{\star}}
\newcommand{\opt}{&&&\optsymbol\hspace*{1mm}}

\newcommand{\eq}[1]{\eqref{eq:#1}}
\newcommand{\eqtwo}[2]{(\ref{eq:#1},\ref{eq:#2})}
\newcommand{\eqthree}[3]{(\ref{eq:#1},\ref{eq:#2},\ref{eq:#3})}
\newcommand{\eqft}[2]{(\ref{eq:#1}--\ref{eq:#2})}
\newcommand{\eqa}[1]{($\rightarrow$\ref{eq:#1})}
\newcommand{\eqatwo}[2]{($\rightarrow$\ref{eq:#1},\ref{eq:#2})}

\newcommand{\bigOsingle}{\mathcal O}
\newcommand{\bigO}[1]{\bigOsingle\left({#1}\right)}
\newcommand{\skewnessSingle}{\mathit{skew}}

{\bfseries}{\itshape}

\newcommand{\ilp}[1]{ILP #1}
\newcommand{\pbs}[1]{PBS #1}
\newcommand{\kurat}{Kuratowski}
\newcommand{\faces}{Facial Walks}
\newcommand{\schnyder}{Schnyder Orders}
\newcommand{\rosen}{Left-Right Coloring}

\newcommand{\mypar}[1]{\smallskip\noindent\textsf{\textbf{#1}}}
\renewcommand{\mypar}[1]{\subparagraph*{#1}}
\newcommand{\aeSection}{\mypar{Algorithm engineering and preliminary benchmarks.} }

\let\oldsum\sum
\renewcommand{\sum}{\oldsum\nolimits}

\definecolor{colorbrewer1}{RGB}{228,26,28}
\definecolor{colorbrewer2}{RGB}{55,126,184}
\definecolor{colorbrewer3}{RGB}{77,175,74}
\definecolor{colorbrewer4}{RGB}{152,78,163}
\definecolor{colorbrewer5}{RGB}{255,127,0}
\definecolor{colorbrewer6}{RGB}{255,255,51}
\definecolor{colorbrewer7}{RGB}{166,86,40}
\definecolor{colorbrewer8}{RGB}{247,129,191}
\definecolor{colorbrewer9}{RGB}{100,100,100}

\tikzset{every picture/.style={mark size=1.5}}

\tikzset{ilp_kurat/.style={colorbrewer1, thick, mark=*, mark size=2pt}}
\tikzset{pbs_kurat/.style={colorbrewer1, thick, mark=*, mark size=2pt, mark options={fill=white}}}
\tikzset{ilp_rosenstiehl/.style={colorbrewer2, thick, mark=triangle*, mark size=3pt}}
\tikzset{pbs_rosenstiehl/.style={colorbrewer2, thick, mark=triangle*, mark size=3pt, mark options={fill=white}}}
\tikzset{ilp_orders/.style={colorbrewer5, thick, mark=square*, mark size=2pt}}
\tikzset{pbs_orders/.style={colorbrewer5, thick, mark=square*, mark size=2pt, mark options={fill=white}}}
\tikzset{ilp_faces/.style={colorbrewer3, thick, mark=diamond*, mark size=3pt}}
\tikzset{pbs_faces/.style={colorbrewer3, thick, mark=diamond*, mark size=3pt, mark options={fill=white}}}
\pgfplotscreateplotcyclelist{alg styles}{ilp_faces,pbs_faces,ilp_orders,pbs_orders,ilp_rosenstiehl,pbs_rosenstiehl,ilp_kurat,pbs_kurat}

\pgfkeys{
 /doplot/.is family, /doplot,
 default/.style = {
  width = \textwidth,
  height = 42.8mm,
  title = TITLE NOT SPECIFIED,
  legend = disable,
  bars = disable,
  axisArgs = {},
  axisType = axis,
  x = COLUMN NOT SPECIFIED,
  barWidth = .7,
  percent = disable
 },
 width/.store in = \doplotWidth,
 height/.store in = \doplotHeight,
 title/.store in = \doplotTitle,
 legend/.store in = \doplotLegend,
 axisArgs/.store in = \doplotAxisArgs,
 axisType/.store in = \doplotAxisType,
 bars/.store in = \doplotBars,
 x/.store in = \doplotXKey,
 percent/.store in = \doplotPercent,
 barWidth/.store in = \doplotYBarWidth
}

\tikzset{/pgf/number format/1000 sep={}}
\pgfplotsset{percentStyle/.style={
 yticklabel={\pgfmathparse{100*\tick}\pgfmathprintnumber{\pgfmathresult}\,\%},
 yticklabel style={/pgf/number format/.cd,fixed,precision=0},}
}

\newcommand{\doplot}[2][]{\pgfkeys{/doplot, default, #1}
\ifthenelse{\equal{\doplotLegend}{disable}}{}{
 \begin{center}
 \def\arraystretch{1.2}
 \begin{tabular}{|rlrl|}
  \hline
  \ref{pgfplots:ilp_kurat}&\ilp\kurat & \ref{pgfplots:pbs_kurat}&\pbs\kurat \\
  \ref{pgfplots:ilp_faces}&\ilp\faces & \ref{pgfplots:pbs_faces}&\pbs\faces\\
  \ref{pgfplots:ilp_orders}&\ilp\schnyder & \ref{pgfplots:pbs_orders}&\pbs\schnyder \\
  \ref{pgfplots:ilp_rosenstiehl}&\ilp\rosen & \ref{pgfplots:pbs_rosenstiehl}&\pbs\rosen\\
  \hline
 \end{tabular}
 \end{center}\vspace*{5mm}
}
\begin{subfigure}{\doplotWidth}
\centering
\begin{tikzpicture}[trim axis left, trim axis right]
 \newcommand{\lex}{}
 \ifthenelse{\equal{\doplotPercent}{disable}}{}{
  \renewcommand{\lex}{percentStyle,}
 }

 \ifthenelse{\equal{\doplotBars}{disable}}{}{
  \let\oldlex\lex
  \renewcommand{\lex}{\oldlex axis y line*=left,}

  \newcommand{\dolabel}{}
  \ifthenelse{\equal{\doplotBars}{nolabel}}{}{
   \renewcommand{\dolabel}{ylabel={instances},}
  }

  \begin{axis}[
    every x tick/.style={black},
    width=\textwidth,
    height=\doplotHeight,
    hide x axis,
    axis y line*=right,
    \dolabel
    ymin=0,
  ] \addplot [
    draw=none,
    ybar,
    bar width=\doplotYBarWidth,
    fill,
    fill opacity=0.2
  ] table [
    col sep=comma,
    x index=0,
    y index=1] {data/#2.csv};
  \end{axis}
 }

 \begin{\doplotAxisType}[
   every x tick/.style={black},
   width=\textwidth,
   height=\doplotHeight,
   \lex
   cycle list name={alg styles},
   \doplotAxisArgs
  ]
  \foreach \i in {ilp_faces,pbs_faces,ilp_orders,pbs_orders,ilp_rosenstiehl,pbs_rosenstiehl,ilp_kurat,pbs_kurat}{
   \addplot table[col sep=comma, x=\doplotXKey, y=\i] {data/#2.csv};
   \ifthenelse{\equal{\doplotLegend}{disable}}{}{
    \label{pgfplots:\i}
   }
  }
 \end{\doplotAxisType}
\end{tikzpicture}
\vspace*{-2.5mm}
\caption{\doplotTitle}
\label{sfg:#2}
\end{subfigure}}

\begin{document}

\maketitle

\begin{abstract}
Given a graph~$G$, the NP-hard \emph{Maximum Planar Subgraph} problem asks for a planar subgraph of~$G$ with the maximum number of edges.
The only known non-trivial exact algorithm utilizes Kuratowski's famous planarity criterion
 and can be formulated as an integer linear program (ILP) or a pseudo-boolean satisfiability problem (PBS).
We examine three alternative characterizations of planarity regarding their applicability to model maximum planar subgraphs.
For each, we consider both ILP and PBS variants, investigate diverse formulation aspects, and evaluate their practical performance.
\end{abstract}

\section{Introduction}
The NP-hard \emph{Maximum Planar Subgraph Problem} (MPS) is a long known problem in graph theory,
 already discussed in the classical textbook by Garey and Johnson~\cite{LiuGeldmacher1979, GareyJohnson1979}.
Given a graph~$G=(V,E)$, we ask for a largest edge subset~$F \subseteq E$ such that the graph induced by $F$ is planar.
The closely related \emph{maximal} planar subgraph problem asks for a set of edges that we cannot
 extend without violating planarity and is trivially solvable in polynomial time.
Sometimes, the inverse measure \emph{skewness} $\skewnessSingle(G)$ is considered, 
 where we ask for the minimum number of edges to delete until obtaining planarity.
MPS has received significant attention for diverse reasons.
Firstly,
 skewness is considered a very natural measure of non-planarity and resides among the most common ones (such as crossing number and genus).
Secondly,
 determining a large planar subgraph is the foundation of the planarization method~\cite{batiniTalamoTamassia1984, chimaniGutwenger2012}
 that is heavily employed in graph drawing:
during planarization, one draws a large---favorably maximum---planar subgraph
 and re-inserts the deleted edges, usually to obtain a low number of overall crossings.
In fact, this gives an approximation algorithm with ratio roughly $\bigO{\Delta \cdot \skewnessSingle(G)}$~\cite{chimaniHlineny17},
 where $\Delta$ denotes the maximum node degree.
Thirdly,
 there are graph problems that become easier when the skewness of the input is small or constant.
E.g., we can compute a maximum flow in time
$\bigOsingle\big( \skewnessSingle(G)^3 \cdot |V| \log |V| \big)$~\cite{hochsteinWeihe2007}, i.e., 
for constant skewness we obtain the same runtime complexity as on planar graphs.

On the positive side, we know that a spanning tree already approximates MPS by $1/3$.
The best known approximation algorithm is due to C\u{a}linescu et al., achieving an approximation ratio of $4 / 9$~\cite{CalinescuFernandesFinklerKarloff1998}.
On the downside, C\u{a}linescu et al.\ also show that the problem is MaxSNP-hard, i.e., there is an upper bound~${\nobreak<1}$ on the obtainable approximation ratio unless $P=\mathit{NP}$.
Just recently, a new algorithm with approximation ratio~$13 / 33$~\cite{Schmid2017} was discovered.
The only non-trivial algorithm in literature for \emph{exactly} computing a maximum planar subgraph is based on
 integer linear programming and Kuratowski's characterization of planarity~\cite{Mutzel1994}. Since its inception over two decades
 ago, \emph{no} other exact algorithm has been proposed, and only few related algorithmic advances improved its performance, see~\cite{ChimaniKleinWiedera2016}.

Besides this famous $K_5$-$K_{3,3}$-subdivision criteria by
 Kuratowski \cite{Kuratowski1930} (see Section~2)
 there is an abundance of planarity criteria.
A (non-complete) list can be found in \cite{LittleSanjith2010,Thomassen1980}.
In this paper,
 we aim at evaluating planarity criteria regarding their usefulness in ILP/PBS formulations to obtain new, alternative exact MPS algorithms.
Naturally, we restrict ourselves to a subset of criteria that we deem promising for this investigation.
We hope to pinpoint new possible ways of considering the problem, to gain new insight into the structure of the MPS, and to lay the groundwork for developing faster exact algorithms.
We present our three new models in Sections~3--5. For each of the possible formulations,
 there are several options and parameter choices.
We report on algorithmic and experimental decisions thereto directly after their description, based on pilot studies%
\footnote{The experimental setting is the same as discussed in Section~\ref{sec:exp}.}.
In Section~\ref{sec:exp} we present a full comparison of the best parameterization for each formulation.

\section{Preliminaries}

In \emph{Linear Programming} (LP),
 one is given a vector $c \in \mathbb R^d$, a set of linear inequalities that define a polyhedron~$P$ in $\mathbb R^d$,
 and asked to find an element $x \in P$ that maximizes $c^\intercal x$.
\emph{Integer Linear Programming} (ILP) additionally requires the components of $x$ to be integral.
Closely related is the concept of \emph{Pseudo Boolean Satisfiability} (PBS),
 sometimes referred to as $0$-$1$-integer linear programming (a special form of ILP: the given polyhedron is a subset of $[0,1]^d$),
 but typically described as a generalization of SAT: its describing constraints are called \emph{clauses} and
 usually have the form of first order Boolean formulae.
 Modern solvers directly support clauses that require a certain number of literals (instead of just one) to be true.
The main difference between PBS and $0$-$1$-ILP is the solution strategy:
 the first uses fast enumeration and clause learning, whereas the latter employs LP-relaxations.
We use these concepts to design models for MPS that can be solved by arbitrary ILP/PBS solvers.

It often is beneficial to not add all constraints to a program but instead identify a relevant subset
 of constraints in the solving process.
This is usually referred to as the \emph{Cutting-Plane Method}.
We utilize it in branch-and-cut-based ILP solving either on fractional or integral solutions.
In PBS solvers, one has to rely on a less sophisticated approach
 that iteratively solves the PBS formula, adds new constraints as appropriate,
 and re-solves the extended formula while maintaining some information from the previous runs.
We refer to clauses that are added iteratively to a PBS formula as \emph{lazy constraints}.

\subsection{Notation}
Throughout this paper, our input graph~$G=(V,E)$ is undirected and simple with $n:=|V|$ and $m:=|E|$.
For general graphs $H$, we refer to its nodes as $V(H)$ and its edges as $E(H)$.
For a directed graph~$H'$ we denote the arcs by $A(H')$ and may write $E(H')$ whenever considering $H'$ undirected.
For any $k \in \mathbb N$, we denote the set $\{0,1,\ldots,k-1\}$ by~$[k]$ and operations on the members are to be understood modulo~$k$.
For any edge~$e$ of $G$, let $\noinci e := V \setminus e$ denote the nodes that are not incident with~$e$.
Given a node~$v$, its neighbors are denoted by~$\neighbors v$.
In a directed graph, we refer to the outgoing (incoming) arcs of a node~$v$ as~$\dout v$ ($\din v$, respectively).
For any two nodes~$u$ and~$v$, we denote an arc from~$u$ to~$v$ by~$uv$.
If unambiguous, we might also refer to an undirected edge~$\{u,v\}$ as~$uv$.
We denote the undirected counterpart of an arc~$a$ as~$\dedge a$.
Given an arc~$a=uv$, we define its reversal~$\crev a:=vu$.
Given a set $X$, the set of all ordered $k$-tuples (all $k$-cardinality subsets)
 consisting of elements from $X$ is referred to as $X^{(k)}$ ($X^{\{k\}}$, respectively).
We abbreviate \emph{pairwise different} by \emph{\pd}

\subsection{Common Foundation of Models}

We assume our input graph $G$ to be biconnected non-planar, with edge weights~$w\colon E(G) \to \mathbb N$ and minimum node degree~3.
This can be achieved in linear time using the \emph{Non-Planar Core} reduction~\cite{ChimaniGutwenger2009} as a preprocessing, without changing the graph's skewness.

All models are presented as ILPs.
Since the PBS counterparts directly map to the ILPs where clauses naturally correspond to constraints, we do not explicitly list the PBS formulations.
We highlight optional constraints that we include in the hope to help quickly finding strong dual bounds with the symbol \optsymbol.
We use solution variables~$s_e \in \{0,1\}$ (for all $e \in E(G)$) that are $1$ if and only if edge~$e$ is in the planar subgraph.
The objective is given by\[
 \max \sum_{e \in E(G)} w(e) \cdot s_e.
\]
We always use Euler's bound on the number of edges in planar graphs:
 \[\sum_{e \in E(G)} s_e \leq 3n-6.\]

\subsection{Known Formulation: Kuratowski Subdivisions}

\begin{theorem}[Kuratowski's Theorem \cite{Kuratowski1930}]
 A graph is planar if and only if it neither contains a subdivision of a $K_5$ nor that of a $K_{3,3}$.
\end{theorem}
Hence, it suffices to ask for any member of the (exponentially sized) set~$\mathcal K(G)$ of all Kuratowski subdivisions that at least one of its edges is deleted:\[
 \sum_{e \in E(K)} s_e \leq |E(K)|-1\quad \forall K \in \mathcal K(G).
\]
This formulation is due to Mutzel~\cite{Mutzel1994}.
Later, Jünger and Mutzel showed that these constraints form facets of the planar subgraph polytope~\cite{JuengerMutzel1996}.
Clearly, we cannot solve the model by writing down every constraint explicitly.
Instead, a sufficiently large but in many practical cases small subset of constraints is identified by a (heuristic) separation procedure.
Over the years, the performance of this approach was improved by
 strong preprocessing~\cite{ChimaniGutwenger2009},
 finding \emph{multiple} violated constraints in linear time~\cite{ChimaniMutzelSchmidt2007},
 and good heuristics~\cite{ChimaniKleinWiedera2016}.

\aeSection
Using an ILP solver, we separate on LP-solutions by rounding the computed fractional values, thus obtaining a graph $H\subset G$
 and extracting Kuratowski subdivisions from $H$. Our experiments indicate that
 rounding down values that are smaller than $0.99$ (and $0.9$ in a second round),
 yields locally optimal (w.r.t.\ the algorithm's parameter space) results.
We use a heap to collect $50$ most violated constraints per LP-solution
 while maintaining linear runtime for the extraction of up to $250$ Kuratowski subdivisions.
For the PBS solver, we iteratively search for satisfying variable assignments and check each for planarity,
 adding up to $50$ lazy Kuratowski constraints each.

\section{Facial Walks}
For any connected planar graph,
 there is an \emph{embedding}~$\Pi$, i.e., a cyclic order of edges around the nodes while the graph
 is drawn planarly.
The regions bounded by the edges are the \emph{faces} of $\Pi$.
The facial walk model is based on an idea developed in \cite{BeyerChimaniHedtkeKotrbcik2016} for computing the genus of a graph; it constitutes
the only known model for the latter problem.
It simulates the face tracing algorithm that visits each face, traversing their borders in clockwise order. Let $\bar f$ be an upper bound on the number of attainable faces.
Let $A$ denote the bidirected counterpart of the undirected edges of~$G$.
We add the following binary variables:
\begin{itemize}
 \newcommand{\ttx}{\tabto{4.1cm}}
 \item $x_i\enskip\forall i \in [\bar f]$
  \ttx Has value $1$ iff face $i$ exists.
 \item $c^i_a\enskip\forall a \in A, i \in [\bar f]$
  \ttx Has value $1$ iff arc~$a$ bounds face~$i$: traversing $i$ in clockiwse\\
  \ttx \enskip order visits $\dedge a$ in the orientation of $a$.
 \item $p^v_{u,w}\enskip\forall v \in V,\ u,w \in \neighbors v$
  \ttx Has value $1$ iff $w$ is the direct successor of $u$ in the cyclic\\
  \ttx \enskip order around $v$.
\end{itemize}
We define the following short-hand notations:
\begin{align*}
p^v(U\times W) &:= \tsum_{u\in U}\tsum_{w\in W} p^v_{u,w},
& x(I) &:= \tsum_{i \in I} x_i,\\
s_v(W) &:= \tsum_{w \in W} s_{vw},
& c^I(J)&:=\tsum_{i\in I}\tsum_{j\in J} c^i_j.
\end{align*}
We then complete our model with the constraints below:
\begin{subequations}
\begin{flalign}
&& n + x([\bar f]) &= 2+\textstyle\sum_{e\in E} s_e \label{eq:fw-euler}\\
&& x_i &= 1 && \forall i \in [3] \opt\label{eq:fw-at-least-3-faces}\\
&& x_i &\geq x_{i+1} && \forall i \in [\bar f - 1]  \opt\label{eq:fw-faces-desc}\\
&& x_i &\leq c^{\{i\}}(A)/3 && \forall i \in [\bar f] \label{eq:fw-at-least-3-darts}\\
&& c_a^i &\leq x_i && \forall a \in A, i \in [\bar f]] \label{eq:fw-face-must-exist}\\
&& c^{[\bar f]}(a) &= s_{\dedge a} && \forall a \in A \label{eq:fw-dart-needs-face}\\
&& c^{\{i\}}(\din v) &= c^{\{i\}}(\dout v) && \forall i \in [\bar f], v \in V \label{eq:fw-in-out}\\
&& c^i_{vw} &\geq c^i_{uv} + p^v_{u,w} - 1 && \forall i \in [\bar f], v \in V, u, w \in \neighbors v \label{eq:fw-succ-1}\\
&& c^i_{uv} &\geq c^i_{vw} + p^v_{u,w} - 1 && \forall i \in [\bar f], v \in V, u, w \in \neighbors v \label{eq:fw-succ-2}\\
&& p^v(u \times \neighbors v) &= s_{vu} && \forall vu \in A \label{eq:fw-succ-exists-1}\\
&& p^v(\neighbors v \times w) &= s_{vw} && \forall vw \in A \label{eq:fw-succ-exists-2}\\
&& p^v(U \times \neighbors v{\setminus} U) &\geq s_v(\{u,\tilde u\}) - 1 && \forall v \in V, \emptyset {\neq} U {\subsetneq} \neighbors v, u \in U, \tilde u \in \neighbors v {\setminus} U \label{eq:fw-cycles}
\end{flalign}
Inequality~\eq{fw-euler} ensures that the number of nodes, faces, and edges satisfy Euler's polyhedron formula.
Constraints~\eq{fw-at-least-3-darts} account for the fact that each face needs at least three arcs.
Conversely, for any arc to be assigned to a face, the face needs to exists \eqa{fw-face-must-exist}.
For any arc whose edge is in the planar subgraph there must exist exactly one face that contains the arc \eqa{fw-dart-needs-face}.
Constraints~\eq{fw-in-out} ensure that the number of inbound arcs equals the number of outbound arcs at a fixed node in a fixed face.
By adding constraints~\eqtwo{fw-succ-1}{fw-succ-2}, we make sure to respect the successor-variables.
Constraints~\eqtwo{fw-succ-exists-1}{fw-succ-exists-2} ensure there are successor variables selected for any edge that is in the solution.
The exponentially large set of cut constraints~\eq{fw-cycles} prohibits multiple cycles in the successor relation.
Optionally, we can force the use of at least the first $3$ faces~\eqa{fw-at-least-3-faces}, otherwise the solution is outerplanar and thus not maximal;
 and we can use faces in order of their indices~\eqa{fw-faces-desc} to break symmetries.

\mypar{Special variables/constraints for degree-3 nodes.}
Consider any degree-3 node~$v$ with neighbors $u^v_0,u^v_1,u^v_2$. If all its incident edges are in the solution, we have two possible cyclic orders.
Otherwise, the cyclic order is even unique.
Thus, instead of introducing six successor-variables~$p^v_{\ldots}$ and constraints~\eqft{fw-succ-1}{fw-cycles}, we can use a single binary variable~$p^v$,
and straight-forwardly simplified constraints, for all $i \in [\bar f]$, $j\in [3]$, and all degree-3 nodes $v$:
\begin{align*}
 c^i_{vu^v_{j+1}} &\geq c^i_{u^v_j v} + (p^v {-} 1) + (s_{v u^v_{j+1}} {-} 1) &
 c^i_{vu^v_{j+2}} &\geq c^i_{u^v_j v} + (p^v {-} 1) + (s_{v u^v_{j+2}} {-} 1) - s_{v u^v_{j+1}}\\
 c^i_{u^v_j v} &\geq c^i_{vu^v_{j+1}} + (p^v {-} 1) + (s_{u^v_j v} {-} 1)&
 c^i_{u^v_j v} &\geq c^i_{vu^v_{j+2}} + (p^v {-} 1) + (s_{u^v_j v} {-} 1) - s_{v u^v_{j+1}}\\
 c^i_{vu^v_j} &\geq c^i_{u^v_{j+1}v} - p^v + (s_{vu^v_j} {-} 1) &
 c^i_{vu^v_j} &\geq c^i_{u^v_{j+2}v} - p^v + (s_{vu^v_j} {-} 1) - s_{v u^v_{j+1}} \\
 c^i_{u^v_{j+1}v} &\geq c^i_{vu^v_j} - p^v + (s_{u^v_{j+1}v} {-} 1) &
 c^i_{u^v_{j+2}v} &\geq c^i_{vu^v_j} - p^v + (s_{vu^v_{j+2}} {-} 1) - s_{v u^v_{j+1}}
\end{align*}
It can be easily verified by a case analysis that the above inequalities cover every possible configuration of neighbors,
 where we might assume that there is at least one neighbor since every maximal solution must be connected.
\end{subequations}
\todo{Compare to genus formulation?}

\aeSection
In our experiments,
the special degree-3 node model did not solve more instances but resulted in a marginal reduction (0.8\%) of runtime; so we use it.
The PBS variant on the other hand suffers from the special degree-3 model, solving 9.38\% less instances.
An ILP variant where we eliminate the solution variables~$s_e$ (directly using the containment variables~$c^i_a$ instead) solved 3.29\% less instances.
We refrain from testing polynomially sized models (betweenness- and index-based instead of constraints~\eqft{fw-succ-1}{fw-cycles}) as
 our exact genus experiments suggest this does not pay off~\cite{BeyerChimaniHedtkeKotrbcik2016}.

\section{Schnyder Orders}

A \emph{partially ordered set} (\emph{poset}) is a pair $P=(S,\prec)$ where $\prec$ is a
strict partial order (transitive, irreflexive, binary relation) over the elements of $S$.
Every poset has a \emph{realizer}, i.e., a set $\mathcal R$ of total orders (transitive, antisymmetric, total, binary relation) on $S$ whose intersection is~$\prec$~\cite{Szpilrajn1930}.
This means that $x\prec y$ if and only if $x <_i y$ for all ${<_i}\in \mathcal R$.
 The \emph{Dushnik-Miller dimension} $\dim P$ of $P$ is the minimum cardinality over all realizers of $P$~\cite{DushnikMiller1941}.
We associate a poset $P_G=(V\cup E, \prec_G)$ to $G$ such that $x \prec_G y$ if and only if $y = \{v,w\} \in E$ and $x \in y$.
The dimension of $G$ is defined as the Dushnik-Miller dimension of $P_G$.
We have

\begin{theorem}[Schnyder's Theorem, 4.1 and 6.2 of \cite{Schnyder1989}]\label{thm:Sch}
 A graph is planar if and only if its dimension is at most three.
\end{theorem}

In fact, a graph with dimension $1$ ($2$) is an isolated node (path, respectively).
Therefore, we propose a model to check for dimension three.
While we could directly use the above criterion for an ILP,
 Schnyder provides another, related and favorable, characterization:
\begin{lemma}[Lemma~2.1 of \cite{Schnyder1989}]\label{lemma:Schnyder1989L2.1}
A graph $G=(V,E)$ has dimension at most $d$ if and only if there exists a set of total orders~${<_1}, \ldots, {<_d}$  on $V$ such that
\begin{enumerate}
\item the intersection of ${<_1}, \ldots, {<_d}$ is empty; and
\item for each edge $\{x,y\}\in E$ and each node $z\notin \{x,y\}$ of $G$, there is at least one order~${<_i}$ such that $x <_i z$ and $y <_i z$.
\end{enumerate}
\end{lemma}

To use this criterion, we add (additionally to $s_e$, $\forall e\in E$) the following binary variables:
\begin{itemize}
 \newcommand{\ttx}{\tabto{55mm}}
 \item $t^i_{u,v} \enskip \forall i \in [3], \forall u,v \in V\colon u \neq v$\quad
  \ttx Has value~$1$ iff $u <_i v$.
 \item $a^i_{e,v} \enskip \forall i \in [3], e \in E, v \in \noinci e$\quad
  \ttx Can have value~$1$ only if $u <_i v\ \forall u \in e$.
\end{itemize}

\noindent We are now able to complete the Schnyder orders ILP by adding:
\begin{subequations}
\begin{flalign}
 && s_e &\leq \tsum_{i=0}^2 a^i_{e,v} && \forall e \in E, v \in \noinci e\label{eq:so-need-order}\\
 && a^i_{e,v} &\leq t^i_{u,v} && \forall i \in [3], e \in E, u \in e, v \in \noinci e\label{eq:so-and-req}\\
 && \tsum_{i=0}^2 t^i_{u,v} &\leq 2 && \forall u,v \in V\colon u \neq v \opt\label{eq:so-intersect}\\
 && t^i_{u,v} + t^i_{v,w} - 1 &\leq t^i_{u,w} && \forall i \in [3],\pd\ u,v,w \in V\label{eq:so-trans}\\
 && t^i_{u,v} + t^i_{v,u} &= 1 && \forall i \in [3], u,v \in V\colon u \neq v \label{eq:so-total}
\intertext{%
Constraints~\eq{so-need-order} ensure that for any edge in the solution the Schnyder-property
 for any non-incident node is satisfied by at least one of the three orders.
By inequalities~\eq{so-and-req}, we make sure that the second requirement of the Schnyder-property is respected.
Transitivity of the total orders is obtained by~\eq{so-trans}.
Finally, we require totality by adding~\eq{so-total}.\newline
\indent 
As Schnyder states~\cite{Schnyder1989}, we may omit the intersection criterion~\eq{so-intersect} as this is satisfied by any non-trivial solution.
Note that for any two adjacent edges~$uv,vw$ in the solution and any $i \in [3]$, we cannot have $a^i_{uv, w} = a^i_{vw, u} = 1$,
 since the orders induced by the $a$-variables are conflicting.
Hence, we might pick a single triangle~$T=\{e_1,e_2,e_3\}$ in the input graph and assign realizing orders to each edge; thereby $v_i$ denotes the node incident to both of $T\setminus\{e_i\}$:}
 && \tsum_{j \in [3]\setminus\{i\}}a^j_{e_i,v_i} &= 0 && \forall i\in[3]\opt\label{eq:so-break-sym}
 \end{flalign}
\end{subequations}
 Analogously, we might apply the same symmetry breaking constraint to two adjacent edges if the graph is triangle-free. (Then $e_3\notin E$, we let $i\in[2]$ but retain the subscript at the sum.)

\aeSection
We tested
 omitting the symmetry breaking constraints~\eq{so-break-sym} (9.12\% less solved instances),
 omitting intersection constraints~\eq{so-intersect} (0.85\% less),
 manually separating the transitivity constraints (which does not change the overall number of solved instances but increases runtime by 4.00\%),
 and using Theorem~\ref{thm:Sch}---the partial order on $V \cup E$---instead (leading to a related but different model that we do not describe here), where we solve 39.89\% fewer instances
 (each when using an ILP solver).
Employing the PBS solver, we obtain similar results for
 omitting symmetry breaking constraints (9.37\% less) and for
 omitting intersection constraints (0.79\% less).
In contrast to above, using lazy transitivity constraints leads to 5.24\% fewer solved instances.
We did not investigate a PBS variant based on Theorem~\ref{thm:Sch} as the ILP performance was already strikingly underwhelming.
We did consider a variant where we use betweenness variables~\cite{CapraraOswaldReineltSchwarzTraversi2011} to describe each of the three total orders.
 This allows us to omit the $a$-variables, but it did not yield satisfactory runtime already on rather trivial instances.
\todo{Test betweenness in our benchmarks?? (We might not have the time for that...)}

\section{Left-Right Edge Coloring}
A \emph{Trémaux tree}~$T$ is a rooted tree in a graph~$H$ such that for any \emph{cotree} edge $\{u,v\} \in E^T_H := E(H) \setminus E(T)$,
 we can traverse the nodes of the tree-path between~$u$ and $v$, such that the levels of the nodes
 (i.e., their distances in~$T$ to the root) are strictly increasing.
\todo{Cite Tŕemaux!}
Any \emph{DFS-tree} (depth-first-search-tree), rooted at the start node, is a Trémaux tree.
For any edge~$e$ we refer to the node closer to the root of~$T$ as~$\csource e$ and the other one as~$\ctarget e$ (this is unique by the Trémaux property).
Any Trémaux tree~$T$ defines a partial order on the nodes:
 for each edge $e \in E(T)$ we set $\csource e \prec \ctarget e$, the partial order is obtained by extending this relationship to its transitive hull.

\newcommand{\gcp}[2]{\ensuremath{#1 \wedge #2}}
\newcommand{\gcpp}[3]{\ensuremath{#1 \wedge #2 \wedge #3}}
\begin{definition}[$T$-alike and $T$-opposite relations]
 We denote the \emph{meet} (closest common ancestor) of two nodes~$u,v$ in $\prec$ by $\gcp uv$.
  De Fraysseix and Rosenstiehl~\cite{FraysseixRosenstiehl1985} define binary relations between cotree edges as follows:
\begin{enumerate}[P1.]
 \item For any $\alpha,\beta,\gamma \in E^T_H$ such that
  $\csource\gamma \prec \csource\alpha \leq \csource\beta \prec \gcpp{\ctarget\alpha}{\ctarget\beta}{\ctarget\gamma} \prec \gcp{\ctarget\alpha}{\ctarget\beta}$,
  $\alpha$ and $\beta$ are \emph{$T$-alike}.
 \item For any $\alpha,\beta,\gamma \in E^T_H$ such that
  $\csource\gamma \prec \csource\alpha \prec \csource\beta \prec \gcpp{\ctarget\alpha}{\ctarget\beta}{\ctarget\gamma} \prec \gcp{\ctarget\beta}{\ctarget\gamma}$,
  $\alpha$ and $\beta$ are \emph{$T$-opposite}.
 \item For any $\alpha,\beta,\gamma,\delta \in E^T_H$ such that
  $\csource\gamma = \csource\delta \prec \csource\alpha = \csource\beta \prec \gcp{\ctarget\alpha}{\ctarget\beta} \prec \gcp{\ctarget\alpha}{\ctarget\gamma}$,
  and $\gcp{\ctarget\alpha}{\ctarget\beta} \prec \gcp{\ctarget\beta}{\ctarget\gamma}$,
 $\alpha$ and $\beta$ are \emph{$T$-opposite}.
\end{enumerate}
\end{definition}

\begin{theorem}[Section~2 of \cite{FraysseixRosenstiehl1985}]
 A connected graph~$H$ with a Trémaux tree~$T$ is planar if and only if there exists a partition of $E^T_H$ into two classes,
  such that any two edges which are $T$-alike ($T$-opposite) belong to the same class (different classes, respectively).
\end{theorem}

Using this characterization, we design a model that describes a Trémaux tree with a feasible bicoloring of cotree edges for any connected, planar subgraph.
We introduce the following set of binary variables, additionally to $s_e$ for all $e\in E$:
\begin{itemize}
 \newcommand{\ttx}{\tabto{2.43cm}}
 \item $t_d\enskip\forall d \in A$
  \ttx Has value $1$ iff arc $d$ is in the Trémaux tree~$T$.
 \item $\ell_{uv}\enskip\forall u,v \in V$
  \ttx Has value $1$ iff node $u$ lies on the path from the root to node $v$ in $T$.\\
  \ttx \enskip Always true for $u = v$ and whenever $u$ is the root of $T$.\\
  \ttx \enskip Models the partial Trémaux ordering $u \prec v \iff \ell_{uv} = 1$.
 \item $r_e\enskip\forall e \in E$
  \ttx Has value $1$ iff edge $e$ is colored red (otherwise colored blue).
\end{itemize}

\begin{subequations}
First, we establish a Trémaux tree.
It has $n-1$ edges~\eqa{lrc-n-darts}, chosen from the planar subgraph~\eqa{lrc-del-darts}.
Its edges seed the partial order on the nodes~\eqa{lrc-seed-order}.
To make sure the order described by the $\ell$-variables is exactly the transitive hull of the tree,
 we need that nodes with the same parent in the tree are not comparable~\eqa{lrc-same-parent-comp}.
Whenever two nodes~$u,v$ are smaller than a third one, $u$ must be comparable to $v$~\eqa{lrc-same-child-comp}.
Constraints~\eq{lrc-triangle}, \eq{lrc-reflex}, and \eq{lrc-order-conflict}
 model transitivity, reflexity, and antisymmetry, respectively.
Finally, the Trémaux tree property---any edge of the planar solution being incident with two comparable nodes---is enforced by constraints~\eq{lrc-tremaux}.
Note that the $t$-variables will always describe a tree, i.e., there are no cycles as this would conflict with the induced partial order by~\eqthree{lrc-seed-order}{lrc-triangle}{lrc-order-conflict}.
\begin{flalign}
 && \textstyle \sum_{d \in A} t_d &= |V|-1 \label{eq:lrc-n-darts}\\
 && t_{d} &\leq s_{\dedge d} && \forall d \in A \label{eq:lrc-del-darts}\\
 && t_d &\leq \ell_d && \forall d \in A \label{eq:lrc-seed-order}\\
 && \ell_{vw} + \ell_{wv} + t_{uv} + t_{uw} &\leq 2 && \forall u \in V, \{uv,uw\} \in \subsetOrder A2 \label{eq:lrc-same-parent-comp}\\
 && \ell_{uw} + \ell_{vw} &\leq 1 + \ell_{uv} +\ell_{vu} && \forall (u,v,w) \in \crossOrder V3 \label{eq:lrc-same-child-comp}\\
 && \ell_{uv} + \ell_{vw} &\leq \ell_{uw} + 1 && \forall (u,v,w) \in \crossOrder V3 \label{eq:lrc-triangle}\\
 && \ell_{uv} + \ell_{vu} &\leq 1 && \forall \{u,v\} \in \subsetOrder V2 \label{eq:lrc-order-conflict}\\
 && \ell_{vv} &= 1 && \forall v \in V \label{eq:lrc-reflex}\\
 && s_e &\leq \ell_{\ctarget e} + \ell_{\csource e} && \forall e \in E \label{eq:lrc-tremaux}\\
\intertext{
Aiming at cutting off some symmetrical solutions,
   we may demand that tree edges and deleted edges are colored blue:
}
 && t_{\csource e\ctarget e} + t_{\ctarget e\csource e} + r_e &\leq 1 && \forall e \in E \opt\label{eq:lrc-tree-blue}\\
 && r_e &\leq s_e && \forall e \in E \opt\label{eq:lrc-del-blue}\\
\intertext{
  We may also enforce a unique Trémaux tree for each given assignment of $s$-variables:
   pick an arbitrary root node~$r \in V$,
   set its incoming arcs to $0$ and those of every other node to~$1$~\eqatwo{lrc-root}{lrc-non-root}.
Let $<_\pi$ denote a fixed non-cyclic order on the adjacency entries for each node.
  We may demand that the first feasible edge in this order is always picked for the tree,
  thus obtaining a distinct feasible DFS-tree for each assignment of $s$-variables~\eqa{lrc-dfs}.}
 && \tsum_{wr \in A} t_{wv} &= 0 && \opt\label{eq:lrc-root}\\
 && \tsum_{wv \in A} t_{wv} &= 1 && \forall v \in V\setminus\{r\} \opt\label{eq:lrc-non-root}\\
 && t_{uw} + \ell_{wv} + s_{uv} &\leq 2 && \forall uv <_\pi uw \in A\label{eq:lrc-dfs}\opt
\end{flalign}
We now establish a feasible bicoloring of the cotree edges.
We define $R_{\alpha,\beta,\gamma}^{u,v}:=C_{\{\alpha,\beta,\gamma\}} + \ell_{\csource \gamma \csource \alpha} + \ell_{uv} - 2$,
 where $C_F := \sum_{d \in F}\big(\ell_d + s_{\dedge d} - t_d - t_{\crev d} - 2\big)$ for any $F\subseteq A$.
\begin{align*}
P^1_{\alpha,\beta}(\gamma,u,v) &:=
 R_{\alpha,\beta,\gamma}^{u,v}
 + \ell_{\csource\alpha \csource\beta} +
 \ell_{\csource\beta u} + \ell_{u\ctarget\gamma} - \ell_{v\ctarget\gamma} + \ell_{v\ctarget\alpha} + \ell_{v\ctarget\beta} - 5,\\
P^2_{\alpha,\beta}(\gamma,u,v) &:=
 R_{\alpha,\beta,\gamma}^{u,v} +
 \ell_{\csource\alpha \csource\beta} +
 \ell_{\csource\beta u} + \ell_{u\ctarget\alpha} - \ell_{v\ctarget\alpha} + \ell_{v\ctarget\beta} + \ell_{v\ctarget\gamma} - 5,\\
\begin{split}
P^3_{\alpha,\beta}(\gamma,\delta,u,v,w) &:=
  R_{\alpha,\beta,\gamma}^{u,v} + C_{\{\delta\}}
  + \ell_{\csource\alpha u} + \ell_{uv} + \ell_{uw}
  + \ell_{u\ctarget\alpha} + \ell_{u\ctarget\beta}\\
  &\phantom{:=} + \ell_{v\ctarget\alpha} - \ell_{v\ctarget\beta} + \ell_{v\ctarget\gamma} - \ell_{v\ctarget\delta}
   - \ell_{w\ctarget\alpha} + \ell_{w\ctarget\beta} - \ell_{w\ctarget\gamma} + \ell_{w\ctarget\delta} - 9.
\end{split}
\end{align*}
We model coloring restrictions of type P1 ($T$-alike),
 P2 ($T$-opposite by one other cotree edge), and P3 ($T$-opposite by two other cotree edges)
 by constraints \eqft{lrc-rosen-i}{lrc-rosen-iii}, respectively:
\begin{flalign}&&&
\begin{array}{@{}l}
   r_{\dedge\alpha} - r_{\dedge\beta} \geq P^1_{\alpha,\beta}(\gamma,u,v)\\
   r_{\dedge\beta} - r_{\dedge\alpha} \geq P^1_{\alpha,\beta}(\gamma,u,v)
 \end{array}
 && \begin{array}{@{}l}
  \forall\text{ arcs } \alpha,\beta,\gamma \in A \text{ of \pd~edges,}\\
  u \neq v \in V\colon \csource\gamma \neq \csource\alpha \land \csource\beta \neq u
  \end{array}
  \label{eq:lrc-rosen-i}&&\\&&&
\begin{array}{@{}l}
  r_{\dedge\alpha} + r_{\dedge\beta} \geq 1 + P^2_{\alpha,\beta}(\gamma,u,v)\\
  r_{\dedge\alpha} + r_{\dedge\beta} \leq 1 - P^2_{\alpha,\beta}(\gamma,u,v)
 \end{array}
 && \begin{array}{@{}l}
  \forall\text{ arcs } \alpha,\beta,\gamma \in A \text{ of \pd~edges,}\\
  u \neq v \in V\colon \csource\gamma \neq \csource\alpha \neq \csource\beta \neq u
  \end{array}
  \label{eq:lrc-rosen-ii}&&\\&&&
\begin{array}{@{}l}
  r_{\dedge\alpha} + r_{\dedge\beta} \geq 1 + P^3_{\alpha,\beta}(\gamma,\delta,u,v,w)\\
  r_{\dedge\alpha} + r_{\dedge\beta} \leq 1 - P^3_{\alpha,\beta}(\gamma,\delta,u,v,w)
 \end{array}
 && \begin{array}{@{}l}
  \forall\text{ arcs } \alpha,\beta,\gamma,\delta \in A \text{ of \pd~edges,}\\
  u,v,w \in V\colon v \neq w \text{ and}\\
  \csource\alpha = \csource\beta \land \csource\gamma = \csource\delta \land \csource\gamma \neq \csource\alpha \neq u\\
  \end{array}&
  \label{eq:lrc-rosen-iii}
\end{flalign}
\end{subequations}%
\begin{figure}
\captionsetup[subfigure]{justification=centering}
\tikzset{every node/.style={draw, circle, minimum size=2mm, inner sep=0, fill=white}}
\tikzset{every edge/.style={draw, ->, thick}}
\tikzstyle{big}=[minimum size=6mm]
\tikzstyle{meta}=[fill=lightgray]
\tikzstyle{label}=[fill opacity=0, draw opacity=0, text opacity=1]
\tikzstyle{vdots}=[draw=none, inner sep=-2mm]
\newcommand{\dotslabel}{\raisebox{2mm}{$\vdots$}}
\centering
\hfill
\begin{subfigure}{0.2\textwidth}
\begin{tikzpicture}
 \path[use as bounding box] (-1.3,-.1) rectangle (1.3,4.5);
 \draw[meta, rounded corners] (-.25, .75) rectangle (.25, 2.25) {};
 \node at (0,0) (cs) {};
 \node at (0,1) (as) {};
 \node at (0,2) (bs) {};
 \node[meta, big] at (-.2,3) (ct) {};
 \node at (0,3) (ctm) {};
 \node[meta, big] at (0,4.2) (abt) {};
 \node at (0,4) (abtS) {};

 \draw (cs) edge node[vdots] {\dotslabel} (as);
 \draw (as) edge node[vdots, fill=lightgray] {\dotslabel} (bs);
 \draw (bs) edge node[vdots] {\dotslabel} (ctm);
 \draw (ctm) edge node[vdots] {\dotslabel} (abt);

 \path (cs) edge[bend left=55] node[label,left,yshift=-1mm] {$\gamma$} (ct);
 \path (bs) edge[bend left=55] node[label,right,yshift=5mm,xshift=-.4mm] {$\beta$} (abt);
 \path (as) edge[bend left=70] node[label,left,yshift=5mm] {$\alpha$} (abt);

 \node[label] at (.5,3.65) (curlyV) {\Huge $\}$};
 \node[label] at (.5,2.65) (curlyU) {\Huge $\}$};

 \node[label] at (.8,2.65) (labelU) {$u$};
 \node[label] at (.85,3.65) (labelV) {$v$};

 \node[label] at (.85,4.3) (labelMeet) {$\gcp{\ctarget\alpha}{\ctarget\beta}$};
 \draw (labelMeet.west) edge[in=90, out=180, thin, -] (abtS);
\end{tikzpicture}
\caption{type P1}
\end{subfigure}\hfill
\begin{subfigure}{0.21\textwidth}
\begin{tikzpicture}
 \path[use as bounding box] (-1.6,-.1) rectangle (1.3,4.5);
 \draw[meta, rounded corners] (-.25, .75) rectangle (.25, 2.25) {};
 \node at (0,0) (cs) {};
 \node at (0,1) (as) {};
 \node at (0,2) (bs) {};
 \node[meta, big] at (-.2,3) (at) {};
 \node at (0,3) (atm) {};
 \node[meta, big] at (0,4.2) (bct) {};
 \node at (0,4) (bctS) {};

 \draw (cs) edge node[vdots] {\dotslabel} (as);
 \draw (as) edge node[vdots, fill=lightgray] {\dotslabel} (bs);
 \draw (bs) edge node[vdots] {\dotslabel} (atm);
 \draw (atm) edge node[vdots] {\dotslabel} (bct);

 \path (cs) edge[bend left=70] node[label,left,yshift=-1mm] {$\gamma$} (bct);
 \path (bs) edge[bend left=55] node[label,right,yshift=5mm,xshift=-.4mm] {$\beta$} (bct);
 \path (as) edge[bend left=55] node[label,left,yshift=5mm] {$\alpha$} (at);

 \node[label] at (.5,3.65) (curlyV) {\Huge $\}$};
 \node[label] at (.5,2.65) (curlyU) {\Huge $\}$};

 \node[label] at (.8,2.65) (labelU) {$u$};
 \node[label] at (.85,3.65) (labelV) {$v$};

 \node[label] at (.85,4.3) (labelMeet) {$\gcp{\ctarget\beta}{\ctarget\gamma}$};
 \draw (labelMeet.west) edge[in=90, out=180, thin, -] (bctS);
\end{tikzpicture}
\caption{type P2}
\end{subfigure}\hfill
\begin{subfigure}{0.33\textwidth}
\begin{tikzpicture}
 \path[use as bounding box] (-2.1,-.1) rectangle (2.4,4.5);
 \node at (0,0) (cds) {};
 \node at (0,1) (abs) {};
 \node at (0,2.5) (t) {};
 \node[meta, big] at (-1.14,4.14) (act) {};
 \node at (-1,4) (actS) {};
 \node[meta, big] at (1.14,4.14) (bdt) {};
 \node at (1,4) (bdtS) {};

 \draw (cds) edge node[vdots] {\dotslabel} (abs);
 \draw (abs) edge node[vdots] {\dotslabel} (t);
 \draw (t) edge node[vdots, rotate=33] {\dotslabel} (actS);
 \draw (t) edge node[vdots, rotate=-33] {\dotslabel} (bdtS);

 \path (cds) edge[bend left=70] node[label, left] {$\gamma$} (act);
 \path (cds) edge[bend right=70] node[label, right] {$\delta$} (bdt);

 \path (abs) edge[bend left=55] node[label, left] {$\alpha$} (act);
 \path (abs) edge[bend right=55] node[label, right] {$\beta$} (bdt);

 \node[label] at (.35,2) (curlyU) {\Huge \scalebox{1}[1.5]{$\}$}};
 \node[label, rotate=213] at (-.95,3.35) (curlyV) {\Huge \scalebox{1}[1.5]{$\}$}};
 \node[label, rotate=-213] at (.47,3.7) (curlyW) {\Huge \scalebox{1}[1.5]{$\}$}};

 \node[label] at (.65,2) (labelU) {$u$};
 \node[label] at (-1.25,3.2) (labelV) {$v$};
 \node[label] at (.15,3.85) (labelW) {$w$};

 \node[label] at (-.1,4.3) (labelMeetAC) {$\gcp{\ctarget\alpha}{\ctarget\gamma}$};
 \draw (labelMeetAC.west) edge[in=90, out=180, thin, -] (actS);

 \node[label] at (2,4.3) (labelMeetBD) {$\gcp{\ctarget\beta}{\ctarget\delta}$};
 \draw (labelMeetBD.west) edge[in=90, out=180, thin, -] (bdtS);
\end{tikzpicture}
\caption{type P3}
\end{subfigure}\hfill\phantom.

\caption{Schematics of configurations inducing $T$-alike and $T$-opposite with ranges to pick nodes~$u,v,w$ from, such that
 constraints~\eqft{lrc-rosen-i}{lrc-rosen-iii} are tight.
 Nodes at the bottom are (closest to) the root.
 Tree paths are straight (cotree edges are bent), partially dotted lines.
 Subgraphs of arbitrary structure (possibly just a single node) are shaded in gray.}
\label{fig:lrc-configs}
\end{figure}
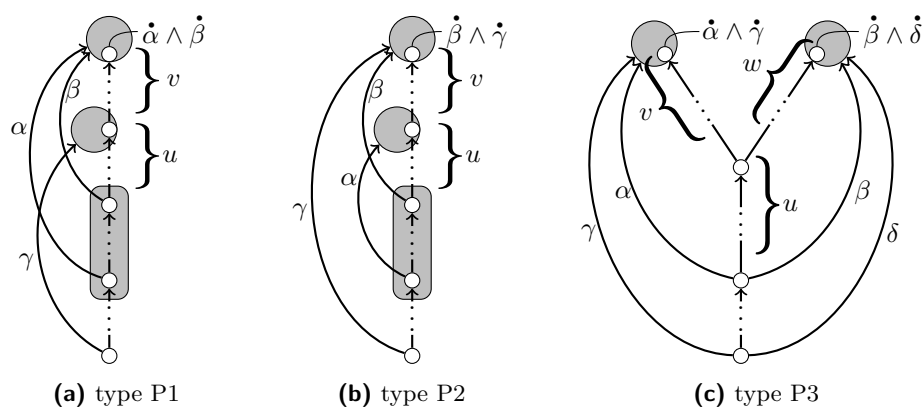%
To comprehend the latter three constraint classes~\eqft{lrc-rosen-i}{lrc-rosen-iii},
 one first needs to understand that for any $F \subseteq A\colon -C_F \in \mathbb N$ by definition (for any feasible variable assignment)
 and $C_F = 0$ if and only if each arc of $F$ is
  a cotree edge of the subgraph induced by the $s$-variables
  and directed from the smaller to the larger node.
Following this pattern, we define the terms~$
 P^1_{\alpha,\beta}(\gamma,u,v),
 P^2_{\alpha,\beta}(\gamma,u,v),
 P^3_{\alpha,\beta}(\gamma,\delta,u,v,w)$
 each equal to~$0$ if and only if we have a configuration of type P1, P2, or P3, respectively, and smaller than or equal to $-1$ otherwise.
Using these terms we can enforce $T$-alike- and $T$-oppositeness for any pair $\alpha,\beta$ as given by constraints~\eqft{lrc-rosen-i}{lrc-rosen-iii};
 see Figure~\ref{fig:lrc-configs} for the selection of nodes $u$, $v$, and $w$.

\mypar{DFS-based branching rule.}
Apart from a traditional automatic selection of branching variables by the ILP solver, we consider a more specialized scheme.
Given a vertex in the branch-and-bound (B\&B) tree, we traverse the locally non-deleted edges of $G$ (i.e., the edges that have a local upper bound of $1$) in their unique DFS order until we find
 an edge~$e$ that is not yet chosen to be in the DFS-tree (i.e., the lower bound of the respective arc variable is not $1$).
We spawn two new B\&B subproblems, where $e$ either is deleted or in the DFS-tree, respectively.
While we lose the potential benefit of always branching on a strongly fractional value, we can fix two instead of just one free variable in both new B\&B branches.

\aeSection
Since we cannot hope to explicitly write down all coloring constraints~\eqft{lrc-rosen-i}{lrc-rosen-iii},
 we separate on integral ILP solutions and use lazy constraints in the PBS variant.
We use a simple $\mathcal O(n^4)$ \todo{check runtime} routine
 that identifies all violated bicoloring constraints for a given non-planar subgraph.
We can terminate this routine prematurely if we consider the set of identified constraints to be locally sufficient.

We evaluated ILP variants where we
 omitted the symmetry breaking constraints~\eqft{lrc-root}{lrc-dfs} (49.91\% less solved instances),
 use our custom branch rule while limiting its application to B\&B-depth at most $6$ (13.22\% more)
  as well as without this limit (32.40\% more),
 and increased the limit of added constraints per LP run from the default of 100 to 1000 (0.93\% more).
Using the PBS solver, we obtain similar results when
 omitting symmetry breaking constraints (65.74\% less).
Furthermore, we investigated a separation routine based on directed cuts\footnote{%
 Directed cut constraints of the form \ $\sum_{w \in W, v \in V \setminus W} t_{wv} \geq 1$ for all $W$ with $\{r\} \subseteq W \subsetneq V.$}
 to cut off infeasible $t$-variable assignments; this does not seem to be beneficial.

\section{Experimental Evaluation}
\label{sec:exp}

\mypar{Setup.}
 All our programs are implemented in \cpp, compiled with GCC 6.3.0, and use the OGDF (version based on snapshot 2017-07-23)~\cite{OGDF}.
 We use SCIP 4.0.1 for solving ILPs with CPLEX 12.7.1 as the underlying LP solver~\cite{SCIP400}.
 For PBS-based algorithms, we utilize Clasp 3.3.3~\cite{CLASP}.
 Each MPS-computation uses a single physical core of a Xeon Gold 6134 CPU (3.2 GHz) with a memory speed of 2666 MHz.
 We apply a time limit of 20 minutes and a memory limit of 8 GB per computation.  \todo{mention hard VS soft limits?}
Our instances and results, giving runtime and skewness (if solved), are available for download at \url{http://tcs.uos.de/research/mps}.

\mypar{Instances and configurations.}
We use the non-planar graphs of the established benchmark sets North~\cite{North1995},
 Rome~\cite[Section 3.2]{DiBattistaGargLiottaTamassiaTassinariVargiu1997}, and a subset of the SteinLib~\cite{KMV00} all of which include real-world instances.
In addition, we generated a set of random regular~\cite{StegerWormald1999} graphs that are \emph{expander graphs} with high probability. In~\cite{ChimaniKleinWiedera2016}
it was observed that such graphs seem to be especially hard at least for the Kuratowski formulation.
For formulations that allow multiple configurations, we determined
the most promising one in a preliminary benchmark on a
  set of 1224 Rome and North graphs, as reported in the previous sections.
 This \emph{fixed} subset of instances was sampled by partitioning the instances
into buckets based on the number of nodes
  and choosing a fixed number of graphs from each bucket with uniform
probability.

For parameters where we had a non-binary choice (e.g., heap size in ILP separation) we rely mostly on the values identified in \cite{hedtkePhd}.

Our algorithms use strong primal heuristics,
 whose common foundation is a maximal planar subgraph algorithm based on the simpler cactus algorithm by C\u{a}linescu et al.,
 with approximation ratio $7 / 18$, that was identified in~\cite[denoted by \texttt{C+}]{ChimaniKleinWiedera2016} to be among the practically best heuristics.

\begin{table}
\centering
\caption{
 Ratios of solved instances. The Kuratowski ILP dominates all other algorithms.
}
\newcolumntype{P}{>{\raggedleft\arraybackslash}p{4.5em}<{\,\%\ }}
\newcommand{\centerMyC}[1]{\multicolumn{1}{c|}{#1}}
\newcommand{\centerMyQ}[1]{\multicolumn{1}{r|}{#1}}
\begin{tabular}{|l|P|P|P|P|}
 \hline
  & \centerMyC{Rome} & \centerMyC{North} & \centerMyC{Expanders} & \centerMyC{SteinLib} \\
 \hline
  \# instances & \centerMyQ{8249} & \centerMyQ{423} & \centerMyQ{480} & \centerMyQ{105} \\
 \hline
 \hline
  \ilp\kurat & \textbf{85.70} & \textbf{73.75} & \textbf{22.75} & \textbf{9.52}\\
 \hline
  \ilp\faces & 17.82 & 29.78 & 4.31 & 2.85\\
 \hline
  \ilp\schnyder & 21.69 & 48.22 & 8.96 & 3.80\\
 \hline
  \ilp\rosen\hspace{1cm} & 36.64 & 60.75 & 12.93 & 3.80\\
 \hline
 \hline
  \pbs\kurat & 77.43 & 69.73 & 10.34 & \textbf{9.52}\\
 \hline
  \pbs\faces & 15.21 & 30.02 & 0.68 & 0.95\\
 \hline
  \pbs\schnyder & 46.24 & 61.93 & 6.89 & 5.71\\
 \hline
  \pbs\rosen & 65.07 & 66.43 & 10.00 & 7.61\\
 \hline
\end{tabular}
\label{tab:solved}
\end{table}

\begin{figure}
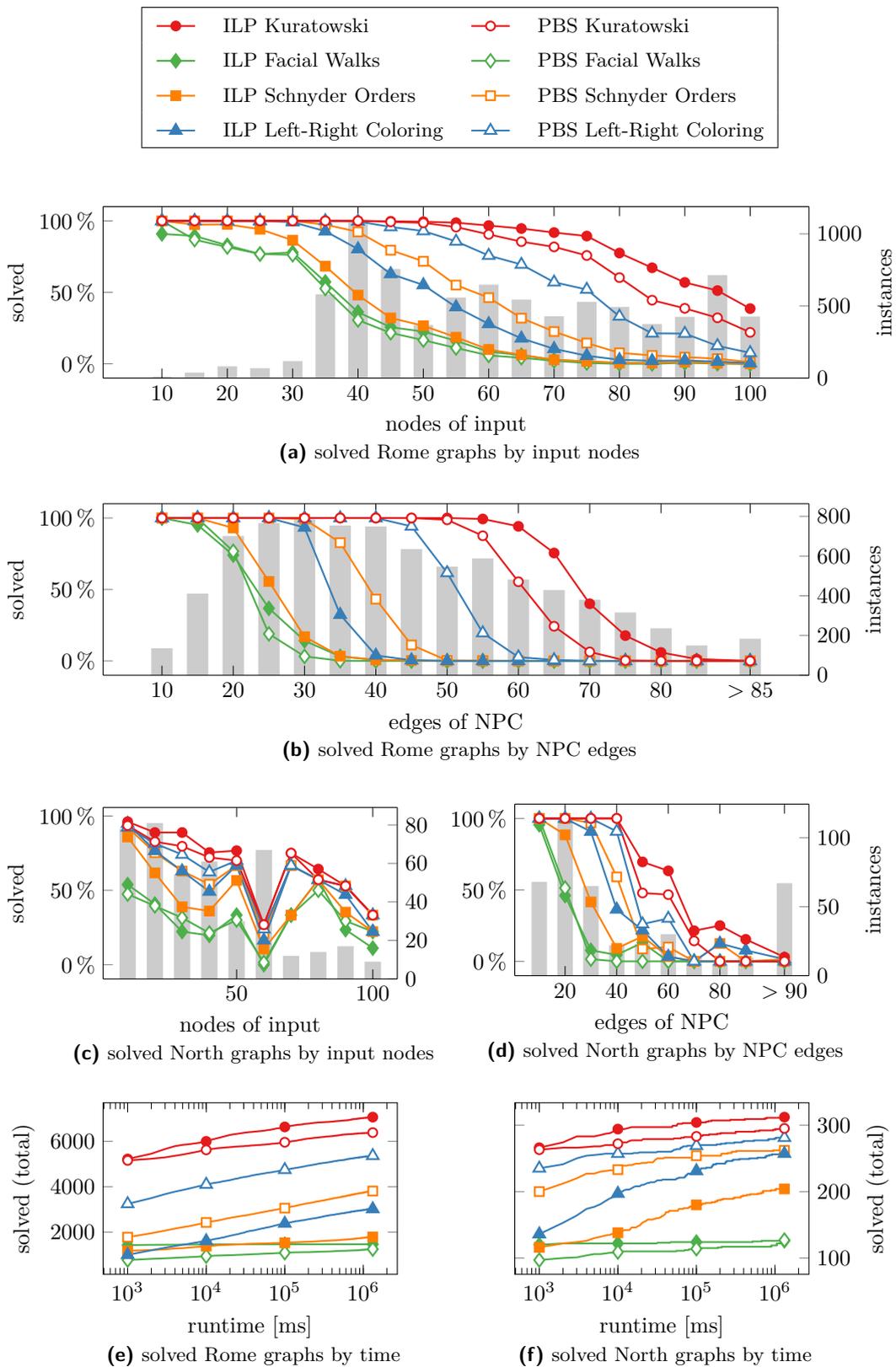

\hspace{6mm}
\begin{minipage}{.9\textwidth}
\centering
\captionsetup[subfigure]{justification=centering}
\newcommand{\vmargin}{\\\vspace{1.5em}}
\doplot[title=solved Rome graphs by input nodes,
 x = nodes,
 percent,
 bars,
 barWidth=3,
 legend,
 axisArgs = {
  xlabel = nodes of input,
  ylabel = solved
 }
]{solved-rome-by-nodes}\vmargin
\doplot[title=solved Rome graphs by NPC edges,
 x = NPCedges,
 percent,
 bars,
 barWidth=3,
 axisArgs = {
  xlabel = edges of NPC,
  ylabel = solved,
  xtick = {10,20,30,40,50,60,70,80,92.5},
  xticklabels = {10,20,30,40,50,60,70,80,$> 85$}
 }
]{solved-rome-by-npc-edges}\vmargin
\doplot[title=solved North graphs by input nodes,
 x = nodes,
 percent,
 barWidth=6,
 bars = nolabel,
 width = .49\textwidth,
 axisArgs = {
  xlabel = nodes of input,
  ylabel = solved
 }
]{solved-north-by-nodes}\hfill
\doplot[title=solved North graphs by NPC edges,
 x = NPCedges,
 percent,
 barWidth=6,
 bars,
 width = .49\textwidth,
 axisArgs = {
  xlabel = edges of NPC,
  ylabel = {},
  xtick = {20,40,60,80,105},
  xticklabels = {20,40,60,80,$> 90$}
 }
]{solved-north-by-npc-edges}\vmargin
\doplot[title=solved Rome graphs by time,
 x = time,
 width = .49\textwidth,
 axisArgs = {
  xlabel = runtime [ms],
  ylabel = solved (total),
  mark indices = {1,10,100,1318}
 },
 axisType=semilogxaxis
]{solved-rome-by-time}\hfill
\doplot[title=solved North graphs by time,
 x = time,
 width = .49\textwidth,
 axisArgs = {
  xlabel = runtime [ms],
  ylabel = solved (total),
  yticklabel pos=right,
  mark indices = {1,10,100,1318}
 },
 axisType=semilogxaxis
]{solved-north-by-time}
\end{minipage}
\caption{Success rate and runtime}
\label{fig:solved}
\end{figure}

\begin{figure}
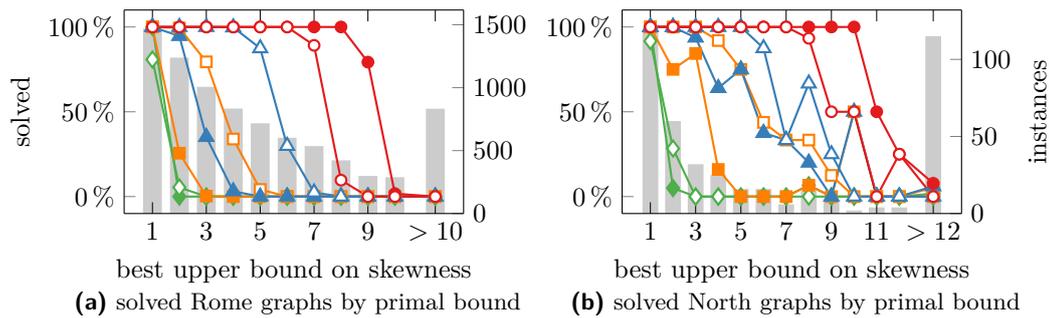

\hspace{8mm}
\begin{minipage}{.9\textwidth}
\captionsetup[subfigure]{justification=centering}
\newcommand{\vmargin}{\\\vspace{1.5em}}
\centering
\doplot[title=solved Rome graphs by primal bound,
 x = skewPrimal,
 percent,
 width=.48\textwidth,
 bars = nolabel,
 axisArgs = {
  xlabel = best upper bound on skewness,
  ylabel = solved,
  xtick = {1,3,5,7,9,11.5},
  xticklabels = {1,3,5,7,9,$>10$}
 }
]{solved-rome-by-skew-ub}\hfill
\doplot[title=solved North graphs by primal bound,
 x = skewPrimal,
 percent,
 width=.48\textwidth,
 bars,
 axisArgs = {
  xlabel = best upper bound on skewness,
  ylabel = {},
  xtick = {1,3,5,7,9,11,13.5},
  xticklabels = {1,3,5,7,9,11,$>12$}
 }
]{solved-north-by-skew-ub}\vmargin
\end{minipage}
\caption{Relation of skewness (bounds) and success rate on Rome graphs. The same legend as in Figure~\ref{fig:solved} applies.}
\label{fig:by-skew}
\end{figure}

\mypar{Results.}
Table~\ref{tab:solved} summarizes the ratios of solved instances.
Evidently, the Kuratowski ILP dominates all implementations.
To our surprise, the rather intricate left-right edge coloring model constitutes the most successful one among the new variants.
The facial walk model falls behind all other formulations.
A similar picture is obtained from a more detailed look at the success rates.
In Figure~\ref{sfg:solved-rome-by-nodes} (\ref{sfg:solved-rome-by-npc-edges}),
 we show the relative number of solved instances among the Rome graphs over the nodes in the input (resp.\ number of edges in the non-planar core),
 clustered to the nearest multiple of five.
As expected, the more edges there are in the core, the harder the instance is in practice.
This is particularly clear on the Rome graphs and becomes a little distorted on the North graphs, see Figures~\ref{sfg:solved-north-by-nodes} and \ref{sfg:solved-north-by-npc-edges}, that include some instances where we have to delete very
 few edges to obtain a (near) triangulation with an (almost) trivial upper bound.
Figures~\ref{sfg:solved-rome-by-time} and \ref{sfg:solved-north-by-time} show the number of solved instances over our total runtime.
The runtime is represented logarithmically.
Again, the Kuratowski ILP is the clear winner and solves more instances than any other variant at any point in time.
While the number of solved instances for all algorithms skyrockets in the first milliseconds and only very slowly increases over the course of $20$ minutes,
 we can see that some algorithms gain more than others from an increase in runtime.
Surprisingly, the Schnyder orders ILP seems to benefit only on the considerably harder North graphs from increasing the runtime.
In most cases, particularly on the Rome and North instances,
 the PBS variant is stronger than its ILP counterpart, with a clear exception for the Kuratowski model.
Finally, Figure~\ref{fig:by-skew} relates upper bounds on the skewness with the number of solved instances.
We can see that there is little success on graphs with a skewness larger than $12$,
 on both the Rome and North set.
 The same holds, although not as clear cut, for the other instance sets.

\section{Findings and Conclusion}

The main goal of this paper was to investigate novel ways of approaching the MPS problem, after over two decades of no progress w.r.t.\ exact models.
We succeeded in the sense that we showed that there are indeed viable alternatives. However, we also showed experimentally that a modern implementation
of the old Kuratowski formulation remains the strongest option to solve MPS in practice. Although negative, this is an interesting
observation.

We should keep in mind that the thereby required efficient separation builds upon years of algorithmic development~\cite{BoyerMyrvold2004, ChimaniMutzelSchmidt2007},
and it is the only ILP where we currently know how to (heuristically) separate on \emph{fractional} solutions.
Equipped with similar tools, i.e.,
 a sensible rounding scheme and a linear time separation routine (a modified left-right planarity test),
 the left-right edge coloring formulation might yield very competitive performance.
 This, in fact, may be a reasonable target for future research.

For the genus problem, a facial walk model similar to our MPS formulation is the only known feasible approach. However,
 we clearly see that it is not favorable for MPS as we have stronger and more direct options at our disposal.
The facial walk model optimizes over all possible embeddings (there are exponentially many already for a fixed subgraph) of all planar subgraphs,
 which might help explain its underwhelming performance.
The Schnyder orders model does not perform very well in practice despite its very elegant characterization.
This might be due to the fact that in contrast to the left-right edge coloring, we search for three feasible orders on the planar subgraph
instead of just one (the partial order corresponding to the Trémaux tree).
To solve the Schnyder orders model efficiently,
 a fast solver for linear ordering problems seems to be required.
The Schnyder and left-right edge coloring PBS formulations usually beat their ILP counterparts, indicating that their LP relaxations are rather weak.
As expected, the expander graphs constitute a particularly hard class of instances and may be a good starting point for tuning and extending our algorithms.

Finally, the strong performance of the Kuratowski model (in particular the ILP variant) is a clear indication
that it deserves more attention in the future. The fact that no additional strong constraint classes have been identified 
for more than two decades is provocative.

\clearpage
\bibliography{lit}

\end{document}